\documentclass[preprint,showpacs,preprintnumbers,amsmath,amssymb,nofootinbib]{revtex4}

\usepackage{etex}

\usepackage{amsthm,amscd,amsbsy,array}
\usepackage{bm}
\usepackage{soul} 

\usepackage{graphics,graphicx,xcolor}

\usepackage{soul} 

\usepackage[colorlinks=true, pdfstartview=FitV, linkcolor=blue, citecolor=blue, urlcolor=blue]{hyperref} 

\newcommand{\gb}{\colorbox{green}}

\newcommand{\dgreen}{\textcolor[rgb]{0,0.35,0}}

\newenvironment{redtext}{\color{red}}{\ignorespacesafterend} 
\newenvironment{bluetext}{\color{blue}}{\ignorespacesafterend} 
\newenvironment{greentext}{\color{green}}{\ignorespacesafterend}
\newenvironment{dgreentext}{\color{\dgreen}}{\ignorespacesafterend}
 
\newcommand{\bblue}{\begin{bluetext}} 
\newcommand{\eblue}{\end{bluetext}} 
\newcommand{\bred}{\begin{redtext}}
\newcommand{\ered}{\end{redtext}}
\newcommand{\bgreen}{\begin{greentext}}
\newcommand{\egreen}{\end{greentext}}
\newcommand{\bdgreen}{\begin{dgreentext}}
\newcommand{\edgreen}{\end{dgreentext}}
\numberwithin{equation}{section}

\let\ssection=\section
\renewcommand{\section}{\setcounter{equation}{0}\ssection}

\newcommand{\cH}{{\mathcal{H}}}
\newcommand{\cQ}{{\mathcal{Q}}}

\newcommand{\cL}{{\mathcal{L}}}

\newcommand{\cM}{\mathcal{M}}

\newcommand{\bp}{{\bf p}}

\newcommand{\bx}{{\bm{x}}}

\newcommand{\bX}{{\mathbf{x}}}

\def\smallover#1/#2{\hbox{$\textstyle\frac{#1}{#2}$}} %

\def\aand{{\quad\text{and}\quad}}

\def\bp{{\bm{p}}}

\def\parag{\hfil\break} 
\def\kikezd{\parag\underbar}

\def\benu{\begin{enumerate}}
\def\eenu{\end{enumerate}}
\def\beq{\begin{equation}}
\def\eeq{\end{equation}}
\def\beqa{\begin{eqnarray}}
\def\eeqa{\end{eqnarray}}

\def\barray{\left(\begin{array}}
\def\earray{\end{array}\right)}
\def\barraynb{\begin{array}}
\def\earraynb{\end{array}}

\def\IR{{\mathbb{R}}} 

\def\?{\quad{\gb{\fbox{\texttt{?}}\;}}\quad}
\def\p{{\partial}}

\def\v0{\mathbf{0}}

\def\beq{\begin{equation}}
\def\eeq{\end{equation}}
\def\bea{\begin{eqnarray}}
\def\eea{\end{eqnarray}}

\def\p{\partial}

\def \p{{\partial}}

\def\6{\partial}
\def\7{\tilde}
\def\8{\widehat}
 \def\bx{{\bf x}}

\def\G11{\Gamma_{11} }

\newcommand{\const}{\mathop{\rm const.}\nolimits}
\newcommand{\half }{\frac{1}{2}}

\def\smallover#1/#2{\hbox{$\textstyle\frac{#1}{#2}$}} %
\def\smallcirc{{\raise 0.5pt \hbox{$\scriptstyle\circ$}}}
\def\2{{\smallover1/2}}

\newcommand{\mybox}[1]{\fbox{$\;\displaystyle{#1}\;$}}

\newcommand{\medbox}[1]{\fbox{%
\rule[-10pt]{0pt}{25pt}$\;\;\displaystyle{#1}\;\;$}%
}

\let\ssection=\section
\renewcommand{\section}{\setcounter{equation}{0}\ssection}

\def\besub{\begin{subequations}}
\def\esub{\end{subequations}}

\begin{document} 

\preprint{\texttt{arXiv:1903.01436v3 [gr-qc] }}

\title{``Kepler Harmonies'' and conformal symmetries
\\[6pt]
}

\author{
P.-M. Zhang${}^{1}$\footnote{e-mail:zhpm@impcas.ac.cn},
M. Cariglia${}^{2}$\footnote
{e-mail: marco.cariglia@ufop.edu.br},
M. Elbistan${}^{1,3}$\footnote{mailto:mahmut.elbistan@lmpt.univ-tours.fr},
G. W. Gibbons${}^{4}$\footnote{mailto:G.W.Gibbons@damtp.cam.ac.uk},
P. A. Horvathy${}^{1,3}$\footnote{mailto:horvathy@lmpt.univ-tours.fr}
}

\affiliation{
${}^1$Institute of Modern Physics, Chinese Academy of Sciences
\\ Lanzhou, China
\\
${}^{2}${\small DEFIS, Universidade Federal de Ouro Preto,  MG-Brasil},
\\
${}^3$ Institut Denis-Poisson CNRS/UMR 7013 - Universit\'e de Tours - Universit\'e d'Orl\'eans Parc de Grammont, 37200, Tours, (France)
\\
${}^4$ D.A.M.T.P., Cambridge University, U.K.
\\
}

\date{\today}

\pacs{\\
45.50.Pk Celestial mechanics,\\
04.20.-q Classical general relativity,\\
04.30.-w Gravitational waves 
}

\begin{abstract}
Kepler's rescaling  becomes, when
``Eisenhart-Duval lifted'' to  $5$-dimensional ``Bargmann'' gravitational wave spacetime, an ordinary spacetime symmetry for motion along null geodesics, which are  the lifts of Keplerian trajectories. The lifted rescaling generates  a well-behaved conserved Noether charge upstairs, which takes an unconventional form when expressed  in conventional terms. This conserved quantity  seems to have escaped attention so far. Applications include the Virial Theorem and also Kepler's Third Law.  
The lifted Kepler rescaling is a Chrono-Projective transformation. The results extend to celestial mechanics and Newtonian Cosmology.\\

\end{abstract}

\maketitle

\tableofcontents

\section{Introduction}\label{Intro}

In today's language, Kepler's Third Law  of planetary motion \cite{Harmonices}
states that the planetary trajectories are taken into themselves under the  rescaling
\beq
t \to \Lambda^3\, t\,,
\quad
\bX \to \Lambda^2\,\bX\,, \quad
\Lambda=\const
\label{K3}
\eeq
where $t$ denotes non-relativistic time and $\bX$ the planet's position.
An intriguing feature is that the standard Lagrangian in $3+1$ non-relativistic dimensions changes under (\ref{K3}) as,
\beq
\label{KLag} 
\cL_{Kepler}=\frac{m}{2}\left(\frac{d{\bX}}{dt}\right)^2+\frac{GmM_{\odot}}{|\bX|}\,
\quad \to \quad
\Lambda^{-2}\, \cL_{Kepler}\,,
\eeq
and is therefore  \emph{not a symmetry} in the sense that the Lagrangian does not change by a total derivative;
some textbooks call it a ``similitude" \cite{LLMech}.

The aim of this Note  is to celebrate the 400 years of Kepler's \textit{``Harmonices Mundi''} which first stated the Third Law \cite{Harmonices} by providing new insight  by ``Eisenhart-Duval lifting''  the  problem to  ``Bargmann'' space, which is in fact the space-time of a plane gravitational wave in 5-dimensions \cite{Eisenhart,Bargmann,DGH91, Brinkmann}. 
 The classical motions downstairs are  the projections from the Bargmann space of \emph{null geodesics}.
 
The clue is then that Kepler's rescaling is  a \emph{Chrono-Projective transformation} \cite{DThese} 
which becomes, when lifted to ``Bargmann space'', a particular type of conformal isometry \cite{5Chrono,DGH91}, which acts as a perfectly well behaving symmetry for null geodesics ``upstairs'' and provides us with a perfectly well behaved
 conserved quantity however when expressed in terms of the original non-relativistic variables ``downstairs'', this quantity takes an unconventional form.

\section{Kepler rescaling}\label{KeplerSec}
 
Let us recall that the Bargmann manifold of a $(d,1)$ dimensional non-relativistic system is  a $d+2$ dimensional manifold $\cM$ endowed with a metric with Lorentz signature which also carries a covariantly constant null vector $\xi$ which we call here the ``vertical vector''. In the case we are interested $d=3$ ; in suitable coordinates $\bX\in\IR^3, t, s\in\IR$, the metric and the vertical vector are,   
\beq
g_{\mu\nu}
dx^{\mu}dx^{\nu} = d\bx^2 + 2dtds +\frac{2GM_{\odot}}{|\bX|}dt^2
\quad\aand\quad
\xi=\p_s\,,
\label{Keplermetric}
\eeq 
respectively. Moreover, $\Delta(\frac{1}{|\bX|})=0$ for $\bX\neq0$ and therefore the  metric 
 is Ricci flat 
\cite{DGH91,Brinkmann} --- it is  a vacuum solution of the  Einstein equations. In other words, it is a \emph{gravitational wave in $5D$}.

The geodesics in Bargmann space are described by the  Lagrangian
\beq
\cL_{geo} = \half (\dot{\bX})^2 +\dot{t}\dot{s}
+\frac{GM_{\odot}}{|\bX|}\,{\dot{t}}^2\,,
\label{KgeoLag}
\eeq
where the ``dot'', $\dot{(\,\,)}=d/d\sigma$ is the derivative w.r.t. an affine parameter $\sigma$. Then (\ref{KgeoLag}) implies the equations of motion
\besub
\begin{align}
\ddot \bX &= -GM_{\odot}{\dot{t}}^2\frac{\bX}{|\bX|^3}\,,
\label{XEOM}
\\
\ddot t &= 0 \,, 
\label{UEOM} 
\\
\frac{\;d}{d\sigma}\big(\dot{s} +\frac{2GM_{\odot}}{\; |\bX|}\, \dot{t}\big)  &=  0 \,.
\label{VEOM} 
\end{align}
\label{EOM}
\esub

The non-relativistic spacetime is identified with the quotient of  $\cM$ by the integral curves of $\xi$; the non-relativistic motions --- in our case the Kepler orbits  --- are the projections to non-relativistic space-time of
 the null geodesic of the $5$ dimensional metric   (\ref{Keplermetric}).

\emph{Chrono-Projective transformations} were
introduced  originally in the Newton-Cartan context  \cite{DThese}.
In Bargman terms they are conformal mapping  of $\cM$,
$f^*g_{\mu\nu}= \Omega^2(t)g_{\mu\nu}$, which leave the \emph{direction} of $\xi$ invariant \cite{5Chrono} \footnote{Transformations which project down are those which strictly preserve the vertical vector, $L_Y\xi = 0$ \cite{Bargmann,DGH91}.} . Working  infinitesimally,
\beq 
L_Yg_{\mu\nu}=2\omega(t)\,g_{\mu\nu}\,
\aand
L_Y\xi = \psi(t)\,\xi\,.
\label{confom}
\eeq
 
Lifting the Kepler rescaling (\ref{K3}) to $5D$ Bargmann space as
\besub
\begin{align} 
&t \to \Lambda^3\, t\,,
\;\;
\bX \to \Lambda^2\, \bX\,,
\;\;
s \to \Lambda\, s
\label{finlift}
\\
&Y=3t\,\p_t+
2\bX\,\p_{\bX}
+s\,\p_s
\label{inflift}
\end{align}
\label{K3lift}
\esub
rescales the metric conformally, 
$  
g_{\mu\nu}dx^{\mu}dx^{\nu} \to \Lambda^4\,g_{\mu\nu}dx^{\mu}dx^{\nu}\,.
$ 
It does not preserve $\xi$, though, only its direction,
\beq
\p_s \to \Lambda^{-1} \p_s\,,
\quad\text{i.e.}\quad L_Y\p_s =-\p_s\,, 
\label{Lambdaxi}
\eeq
and is therefore Chrono-Projective. 
The geodesic Lagrangian (\ref{KgeoLag})
 scales, under the lifted Kepler scaling (\ref{K3lift}),
 as
$ 
\cL_{geo} \to  \Lambda^4 \cL_{geo}\,.
$ 
At the first sight, this  appears to be a no better behaviour than
 ``downstairs'', (\ref{KLag}), -- and this is indeed so when the
geodesic is \emph{timelike or spacelike}. \emph{However for null geodesics the Lagrangian is constrained to vanish},
\beq
\cL_{geo}=0,
\label{nullLag}
\eeq
which makes it invariant~: \emph{lifted Kepler rescalings act, for null geodesics, as symmetries}. 

Let us emphasise that ``upstairs'' i.e., in the $5D$  gravitational wave space-time, no additional constraint  is required; Noether's theorem works for any conformal vectorfield which leaves the geodesic Lagrangian invariant. The associated conserved quantity for \emph{ motion along a null geodesic in $5D$} is, in our case, 
\beq \medbox{
\cQ = 3 t p_t + 2 x^i p_i  + s p_s \, ,
}
\label{KeplerQ}
\eeq 
whose conservation can also be checked by a direct calculation~: in terms of the canonical momenta
 the eqns of motion (\ref{EOM}) imply that 
$ 
p_t = \dot{s} +\frac{2GM_{\odot}}{|\bX|}\, \dot{t}
$
and $p_s =\dot{t}  
$
 are constants of the motion. Then deriving $\cQ$ and using the eqns of motion we get 
\beq
\dot{\cQ}=4\cL_{geo}=0\,,
\label{dotQL}
\eeq
 because, precisely, our geodesics are {null}.
Conversely, the generating vector field $Y$ in (\ref{inflift}) is recovered as 
$ 
Y^\mu = \Big\{x^\mu,\cQ\Big\}={\p \cQ}/{\p p_\mu}\,.
$ 

The geodesic Hamiltonian is
\beq
\cH_{geo}= \half {p_i p_i} -\frac{G M_{\odot}}{|\bX|}p_s^2+p_tp_s.
\eeq
Performing a Legendre transformation, this Hamiltonian  becomes the geodesic Lagrangian, $\cH_{geo}=\cL_{geo}$. Moreover, identifying $p_s$ with $m$ the mass in $3D$ and expressing $p_t$ from the null condition $\cH_{geo}=0$ yields 
\beq
p_t = - \left(\frac{1}{2m} p_i p_i-\frac{GmM_{\odot}}{|\bX|}\right) = -E\,,
\label{PUE}
\eeq
which is (minus) the non-relativistic energy.
Then from $\dot{t}=p_s=m$ we infer that  $d/d\sigma = m d/dt$. 
Denoting
$d/dt$ by ``prime", $p_i=m (x^i)'$ and
putting the geodesic Lagrangian equal to zero yields
\beq
{s}'(t) = - \left(\frac{1}{2}(\bX')^2 + \frac{GM_{\odot}}{|\bX|}\right)=-
\frac{1}{m} \cL_{Kepler}\,.
\label{Vprime}
\eeq
The change of $s$ along a lightlike geodesic is thus proportional to \emph{minus the 3D Kepler action calculated along the projected trajectory} \cite{Brinkmann}\,,
\beq
s(t) = \frac{\cQ}{m}+ 3t\frac{E}{m}-
\frac{2 (p_ix^i)}{m} 
= s_0-
\frac{1}{m}\int_0^t \!\cL_{Kepler}(\bX(\tau),\bX'(\tau)) \,d\tau\,.
 \label{VchangeK3}
\eeq
The conserved quantity (\ref{KeplerQ}) can be expressed ``mostly'' but \emph{not completely} in $3D$ terms, 
\beq 
\cQ=
 - 3 t E +m \frac{\;\, d}{dt}(x_ix^i) - 
\int_0^t \!\!\cL_{Kepler}(\bX(\tau),\bX'(\tau)) \,d\tau\, 
+ms_0\,,
\label{QKeplerbis}
\eeq 
which explains also why it does not project  down~: it depends on $s_0$. However \emph{subtracting $ms_0$}, we get $Q_{Kepler}=\cQ-ms_0$
\beqa 
\mybox{
Q_{Kepler} 
=
- 3 t E + m \frac{\;\, d}{dt}(x_ix^i) - 
\int_0^t \!\!\!\cL_{Kepler}(\bx(\tau),\bx'(\tau)) \,d\tau
}\quad
\label{projCQ}
\eeqa
where the integration is along the classical trajectory in $3$-space. $Q_{Kepler}$ \emph{is} well-defined and also conserved, as proved along the same lines as for (\ref{dotQL})). We mention that the same expression can also be derived from the original Kepler Lagrangian using a generalization of the classical Noether theorem \cite{Kosinskietal}.

Let us stress that (\ref{projCQ}) is a \emph{local quantity} despite its surprising form, because the classical trajectory  (apart at caustic singularities) is uniquely deteremined by its end points. The integral is just Hamilton's action function.

For $t=0$ both the first and the last terms in (\ref{projCQ}) vanish, leaving us with
$ 
Q_{Kepler} = 2p_i(0)x^i(0)\,.
$ 
Let us record for further reference that $\displaystyle\int_0^t\! \cL_{Kepler}\, dt = \displaystyle\int_0^t\! p_idx^i - E t$, which allows us to rewrite (\ref{projCQ}) also as
\beq
Q_{Kepler}=- 2 t E + m \frac{\;\, d}{dt}(x_ix^i) - 
\int_0^t \!\!\!p_i \frac{dx^i}{dt} \,dt\, .
\label{projCQtris}
\eeq

\section{Applications of our conserved quantity}\label{appliSec}

We restrict our attention henceforth to elliptic motions in the $x_3=0$ plane with $E<0$ and draw some interesting consequences of the conservation of (\ref{projCQ}). Parabolic and hyperbolic motions behave  similarly.

If time is measured so that for $t=0$ we are in the closest (perihelion) position then $Q_{Kepler}$ is just zero. But then  a full period later i.e. for $t=T$, we are back where we started from, so that $\big(d(x_i x^i)/dt\big)(t=T)=0$, yielding,
\beq
 2 T E  + \int_0^T \!\!\!p_i \frac{dx^i}{dt} \,dt=0\, .
\label{preK3}
\eeq
The first consequence is that
expressing the momenta in terms of velocities  allows us to infer the \emph{Virial Theorem}~: \emph{the energy is minus the average kinetic energy for a full period}, 
\beq 
E = - \frac{1}{T} \int_0^T \frac{m}{2} (\frac{d\bx}{dt})^2 dt = -\overline{E}_k \,.
\label{Eaverage} 
\eeq  

The integral in (\ref{preK3}) can actually be determined. Consider the radius vectors drawn from the
focus  where the Sun sits \emph{and also from the other focus} to the current position of the planet. The rates of change of the areas swept out the by these two radius vectors are
\besub
\begin{align}
\frac{dA}{dt} &= \frac{1}{2m}p_\phi \,,    
\label{K2}
\\[6pt]
\frac{dA^\prime}{dt} 
  &=  \frac{b^2}{2p_\phi} \bigl(p_r \frac{dr}{dt}+ p_\phi
\frac{d\phi}{dt}\bigr)\,. 
\label{Tait} 
\end{align}
\label{AA'}
\esub
where $p_\phi$ is the angular momentum.
The first of these relations is  {Kepler's Second Law}, while (\ref{Tait}) might reasonably be called \emph{Tait's Law} \cite{Tait,TaitSteele}; see \cite{GEllis2} for a new, geometric proof. Then for a full period $T$ both radius vectors sweep through the ellipse, and therefore,
\beq
\pi ab=\left\{
\barraynb{lll}
\displaystyle\int_0^T\frac{dA}{dt}dt &=&\displaystyle\frac{p_\phi}{2m}\,T\,,
\\[20pt]
\displaystyle \int_0^T\frac{dA^\prime}{dt}dt
 &=&\displaystyle\frac{b^2}{2p_\phi}
\int_0^T\!\bigl(p_r\frac{dr}{dt}+p_\phi\frac{d\phi}{dt}\bigr)dt\,,
\earraynb\right.
\eeq
where $a$ and $b$ are the semi-major and the semi-minor axes, respectively. From here we infer
\beq
\int_0^T\!\bigl(p_r\frac{dr}{dt}+p_\phi\frac{d\phi}{dt}\bigr)dt=\frac{4\pi^2a^2m}{T}\,.
\eeq
Reinserting this into (\ref{preK3}) and using $E=-GmM_{\odot}/2a$, we end up with the Third Law,
\beq
\frac{a^3}{T^2}=\frac{GM_{\odot}}{4\pi^2}\,.
\label{OK3}
\eeq

Let us observe finally that while the planet goes around once the vertical coordinate changes, by (\ref{VchangeK3}), by
\beq
\Delta s = - \frac{1}{m}\int_0^T\!\!\cL_{Kepler} dt = - \frac{1}{m}
\int_0^T\!\bigl(p_r\frac{dr}{dt}+p_\phi\frac{d\phi}{dt}\bigr)dt+ \frac{E}{m}T =-\frac{4\pi^2a^2}{T}-\frac{GM_{\odot}T}{2a}. 
\label{Deltas}
\eeq

\begin{figure}
\includegraphics[scale=.43]{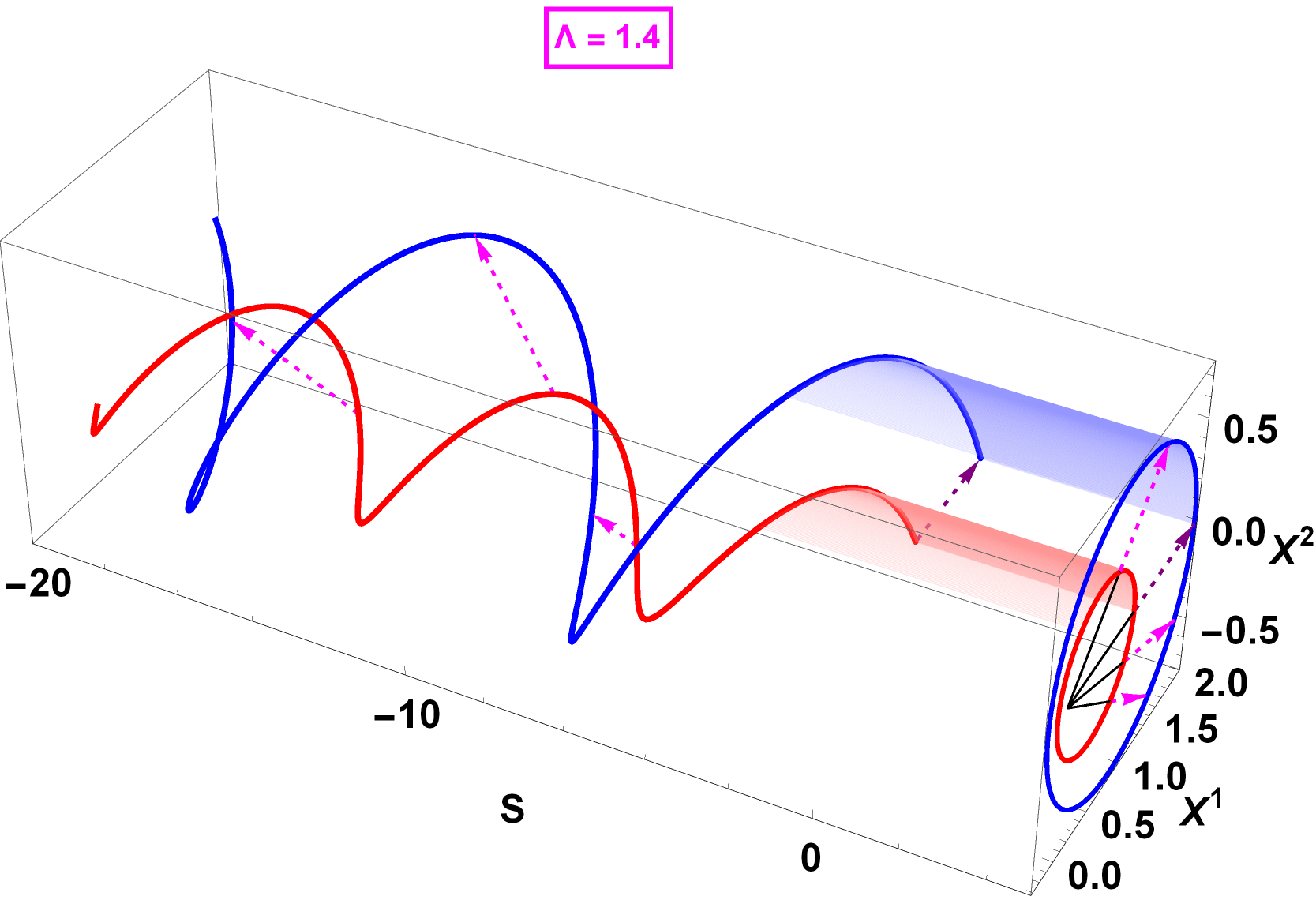}\vskip-3mm
\caption{\textit{\small 
The closed Keplerian trajectories become spirals when lifted to Bargmann space. They are permuted by the lift (\ref{K3lift}) of Kepler's rescaling (\ref{K3}) indicated by arrows. 
}}
\label{KeplerVLambda}
\end{figure}

The equations of motion (\ref{EOM}) can be solved numerically ; it confirms that (\ref{KeplerQ}) is indeed conserved, and also the formulae of this section. The solutions are shown in Fig.\ref{KeplerVLambda}.

\section{Generalization to $N$ bodies}

The Kepler's scaling property holds in fact for all of Newtonian Cosmology \cite{GEllis,GEllis2}.
The $N$-body equations (No sum on $a$),
\beq
m_a\frac{d^2\bx_a}{dt^2}=-\sum_{b\neq a}m_am_b\frac{\bx_a-\bx_b}{\,|\bx_a-\bx_b|^3},
\qquad
 a=1,2,\dots, N
\label{Nbodyeqn}
\eeq
 correspond, in the Bargmann framework, to 
the projections to the $N$-particle configuration space of the null geodesics of the $3N+2$ dimensional metric  \cite{DGH91},
\besub
\begin{align}
&g_{\mu\nu}dx^{\mu}dx^{\nu}=
\sum_{a=1}^N\frac{m_a}{m}d\bx_a^2+2dtds-\frac{2U}{m}dt^2
\,,
\label{Nmetric}
\\[6pt]
&m=\sum_{a}m_a\,,
\qquad
U=-\half\sum_{a,b\neq a}\frac{Gm_am_b}{|\bx_a-\bx_b|}
\,.
\label{NNewtonpot}
\end{align}
\esub
Then the Kepler rescaling (\ref{K3lift}),
$t \to \Lambda^3 t,\,
\bx_a \to \Lambda^2\, \bx_a,\, s \to \Lambda\, s$,
acts plainly conformally, $g_{\mu\nu}dx^{\mu}dx^{\nu}\to \Lambda^4g_{\mu\nu}dx^{\mu}dx^{\nu}$, generating  a symmetry and a conserved charge for null geodesics,\vskip-7mm
\besub
\begin{align}
&\cQ = -3TE+ 2 \sum_a\bx_a\cdot\bp_a  + s m \, ,
\qquad
E=\sum_a \frac{\bp_a^2}{2m_a}+U\,,
\label{NKeplerQ}
\\
&
s = s_0-\frac{1}{m}\int_0^t\!\!\cL_{N}d\tau\,,
\qquad\qquad\qquad\;\;
\cL_{N}=\sum_a \frac{m_a(\bx_a')^2}{2}-U\,.
\label{Naction}
\end{align}
\label{Ncharge}
\esub
Here $\cL_{N}$ is the $N$-body Lagrangian. This charge projects again to a conserved charge of  unconventional form.

\goodbreak
\section{Conclusion}\label{Concl}

This year we celebrate the 400'th
anniversary of Kepler's discovery of his Third Law of planetary motion, which concerns the period and size of geometrically similar bound orbits  \cite{Harmonices}.

Famously, Newton derived this and other properties of the orbits from his Universal Inverse Square Law of Gravitation. This is what we find in most textbooks, e.g.  in \cite{LLMech}.  Since his time there have been many investigations of the geometry and symmetry of these orbits, but none has derived Kepler's Third Law using the  methods introduced by Emmy Noether.

In this paper we start with a previously developed but not well known general formalism called the Bargmann framework of Eisenhart \cite{Eisenhart}, and of Duval et al \cite{Bargmann,DGH91}. It states that the motion may be regarded as the projection of the  motion of light rays moving in a five-dimensional extended spacetime and obtain for the first time Kepler' law as a consequence of Emmy Noether's theorem. 

In detail, lifting Kepler's rescaling, (\ref{K3}), to $5D$ Bargmann space as (\ref{K3lift}) generates there  a well-behaved conserved charge, (\ref{KeplerQ}), for null geodesics. It allows us to integrate the  ``vertical'' motion once the Kepler motions had been determined. 
Subtracting a constant term yields a conserved charge for ordinary planetary motion  of an \emph{unconventional form} \footnote{    
Unconventional conserved quantities for geodesic motion in a curved space were studied recently in \cite{Dimakis}. }.
Rather  incredibly, (\ref{projCQ}) appears to be a \emph{new} conserved quantity which seems to have escaped attention so far. It is in fact \emph{different} from the familiar Runge-Lenz vector as can be understood by recalling their origin: while (\ref{KeplerQ}) is a \emph{scalar} generated by a \emph{conformal Killing vector} of 5D Bargmann space, the  components of the Runge-Lenz vector are associated with 3 \emph{Killing tensors} \cite{DGH91}.

The conserved quantity (\ref{projCQ}) allows us to derive the Virial Theorem, (\ref{Eaverage}), the usual form of Kepler's Third law, (\ref{OK3}); the evolution of the $s$-coordinate is consistent with  Fig.\ref{KeplerVLambda}. 

One can inquire if the Kepler problem admits further spacetime symmetries. The answer is \emph{no}: the  intrinsically defined  Newton-Cartan structure allows for a $5$-parameter
Chrono-Projective group only, composed by rotations, time translations and the Kepler rescaling \cite{DThese}.  For further details and applications of conformal symmetries for gravitational waves, see \cite{AndrPrenc,Conf4GW}.
Other examples of Chrono-Projective transformations include the Schr\"odinger-Newton equations \cite{DuLaz}, hydrodynamics \cite{hydro}, Schr\"odinger  operators \cite{Gundry}  and projective dynamics \cite{Cariglia15}.

The expression $\displaystyle\int_0^T \!\!\!p_i \frac{dx^i}{dt} \,dt$ we used repeatedly in  sec.\ref{appliSec} had actually played a prominent  r\^ole 
in the \emph{Old Quantum Theory}, namely in
the  Nicholson-Bohr-Wilson-Sommerfeld quantization conditions  
  \cite{Nicholson,Bohr,Wilson,Sommerfeld,Wilson2}~: if a coordinate $q$ varies periodically with time, then the quantity  
\beq
\frac{1}{2 \pi h} \oint pdq \,, 
\eeq
where $h= 2 \pi \hbar$ is Planck's constant, should be an \emph{integer}.
For the closed Keplerian orbits we have two such coordinates, $r$ and $\phi$. 
We have in particular the quantization of angular momentum first suggested by Nicholson \cite{Nicholson} and its generalization proposed, independently, by Bohr  \cite{Bohr}, Wilson \cite{Wilson,Wilson2} and by Sommerfeld \cite{Sommerfeld},
\beq
\frac{1}{2 \pi h} \oint p_\phi d\phi = l\,,
\aand 
\frac{1}{2\pi{h}} \oint(p_r dr + p_\phi d\phi) = n\,,
\eeq
where $l$ the total angular momentum and $n$ is the principal quantum number, respectively. The geometrical significance of these relations is given by (\ref{Tait}).

In our study we were helped by 
 that, because of the \emph{Equivalence Principle}, the Keplerian trajectories are independent of the mass.  However, it is illuminating to consider the 3D transformations inherited from those in  5D phase space and  generated by \eqref{KeplerQ} by complementing \eqref{K3}  with  
$
m \rightarrow \Lambda^{-1} m \, . 
$ 
 Details will be discussed elsewhere.

We conclude with the remark that Kepler's game changing
  three laws remain as relevant to our exploration of the universe
  and the laws that govern it today as when they were first formulated.
  No better illustration of this fact may be found than in \cite{Gillessen:2008qv,Takekawa}. Studying the motion of matter around a black hole could provide a test for the validity/corrections of Kepler's laws at the large scale.
  
 \kikezd{Note added}. After our paper was submitted, we came across \cite{Keplar}  which has a vague relation to our work here.
 
\begin{acknowledgments} 
 This paper is dedicated to the memory of Christian Duval (1947-2018). Discussions are acknowledged to Janos Balog, Nathalie Deruelle and  Piotr Kosi\'nski. 
 ME thanks the \emph{Institute of Modern Physics} of the Chinese Academy of Sciences in Lanzhou and  the \emph{Denis Poisson Institute of Orl\'eans-Tours University}.
PH  thanks the \emph{Institute of Modern Physics} of the Chinese Academy of Sciences in Lanzhou  for hospitality. This work was partially supported by the Chinese Academy of Sciences President's International Fellowship Initiative (No. 2017PM0045), and by the National Natural Science Foundation of China (Grant No. 11575254). MC acknowledges CNPq support from projects (303923/2015-6) and (306591/2018-9). 
\end{acknowledgments}
\goodbreak
%


\end{document}